\documentclass[12pt]{article}

\begin{document}
\begin{center}
{\bf On generalized Born$-$Infeld electrodynamics }\\
\vspace{5mm}
 S. I. Kruglov
 \\

\vspace{5mm}
\textit{University of Toronto at Scarborough,\\ Physical and Environmental Sciences Department, \\
1265 Military Trail, Toronto, Ontario, Canada M1C 1A4}
\end{center}

\begin{abstract}

The generalized Born$-$Infeld electrodynamics with two parameters is
investigated. In this model the propagation of a linearly
polarized laser beam in the external transverse magnetic field is
considered. It was shown that there is the effect of vacuum
birefringence, and we evaluate induced ellipticity.
The upper bounds on the combination
of parameters introduced from the experimental data of BRST
and PVLAS Collaborations were obtained. When two parameters are equal to each other, we arrive at Born$-$Infeld
electrodynamics and the effect of vacuum birefringence vanishes. We find the
canonical and symmetrical Belinfante energy-momentum tensors. The
trace of the energy-momentum tensor is not zero and the dilatation
symmetry is broken. The four-divergence of the dilatation current
is equal to the trace of the Belinfante energy-momentum tensor and is
proportional to the parameter (with the dimension of the field strength) of the model.
The dual symmetry is also broken in the model considered.

\end{abstract}

\section{Introduction}

A renewal of interest in the Born$-$Infeld (BI) theory is due to the
development of string theories. Thus, the low energy D-brane
dynamics is governed by a BI type action \cite{Tseytlin}. Firstly,
a non-linear BI electrodynamics was formulated \cite{Born},
\cite{Infeld} to have a finite electromagnetic energy of a point
charge contrarily to Maxwell's electrodynamics. The
fundamental constant $B$ introduced in BI electrodynamics, with the dimension
of the field strength, characterizes the upper bound for the possible electromagnetic fields. When
the $B$ approaches to infinity the BI theory converts to Maxwell's
electrodynamics. Some aspects of wave propagation in Born-Infeld
electrodynamics were investigated in \cite{Plebanski},
\cite{Boillat}, \cite{Gibbons}, \cite{Aiello}.

In this paper, we formulate and investigate the generalized BI
electrodynamics with two parameters. If two parameters introduced are equal,
we come to BI electrodynamics.

The paper is organized as follows. In section 2, the Lagrangian of the model
is motivated. The field equations in generalized BI electrodynamics are formulated
in section 3. We investigate the propagation of a linearly polarized laser beam
in the external transverse magnetic field and evaluate induced
ellipticity in section 4. The upper bounds on the combination
of parameters $\beta$ and $\gamma$ introduced are obtained from the experimental data of BRST
and PVLAS Collaborations. In section 5., we find the canonical and
symmetrical Belinfante energy-momentum tensors, the dilatation
current, and its non-zero divergence. The results are discussed in section 6.

The Heaviside$-$Lorentz system with $\hbar =c=1$ is chosen, and Euclidian metric is
used, $x_\mu=(x_m,ix_0)$, $x_0$ is a time. Greek
letters range from $1$ to $4$ and Latin letters range from $1$ to
$3$.

\section{The model}

Let us consider the Born$-$Infeld Lagrangian density \cite{Born},
\cite{Infeld}
\begin{equation}
{\cal L^{B-I}} =
B^2\left(1-\sqrt{1+2B^{-2}S-B^{-4}P^2}\right),
 \label{1}
\end{equation}
where two Lorentz-invariants are given by
\begin{equation}
S=\frac{1}{4}F_{\mu\nu}^2,~~~~P=\frac{1}{4}F_{\mu\nu}\widetilde{F}_{\mu\nu},
\label{2}
\end{equation}
$F_{\mu\nu}$ is the field strength, and
$\widetilde{F}_{\mu\nu}=(1/2)\varepsilon_{\mu\nu\alpha\beta}F_{\alpha\beta}$
is its dual tensor ($\varepsilon_{1234}=-i$. The parameter $B$ characterizes the upper bound on the possible
field strength. The Lagrangian density (1) is converted to the
Maxwell's Lagrangian density when the fields are weak compared to
the parameter $B$. Thus, for small values of $B^{-2}S$, $B^{-4}P^2$, in leading
order, equation (1) becomes
\begin{equation}
{\cal L^{B-I}}\simeq-S+\frac{1}{2B^2}S^2+\frac{1}{2B^2}P^2.
\label{3}
\end{equation}
At the same time quantum one-loop corrections in QED lead to the Heisenberg$-$Euler
Lagrangian density of self-interaction \cite{Heisenberg}, \cite{Schwinger}, \cite{Adler}
\begin{equation}
{\cal L^{H-E}}= 4aS^2+bP^2,
\label{4}
\end{equation}
where
\begin{equation}
a=\frac{2\alpha^2}{45m_e^4},~~~~b=\frac{14\alpha ^2}{45m_e^4}, \label{5}
\end{equation}
and $\alpha =e^2/(4\pi )\simeq 1/137$ is the fine structure
constant, and the $m_e$ is the electron mass.
We imply that in quantum BI theory the same quantum corrections appear as in QED at least in the perturbation theory because at large value of $B$ the BI Lagrangian becomes the classical electrodynamics Lagrangian. As a result, we may consider the approximate BI Lagrangian density (3) with quantum corrections (4) that leads to
the effective Lagrangian density:
\begin{equation}
{\cal L}_{eff}={\cal L^{B-I}}+{\cal L^{H-E}}\simeq-S+\frac{1}{2\beta^2}S^2+\frac{1}{2\gamma^2}P^2,
\label{6}
\end{equation}
where
\begin{equation}
\frac{1}{\beta^2}=\frac{1}{B^2}+8a,~~~~~~\frac{1}{\gamma^2}=\frac{1}{B^2}+2b.
\label{7}
\end{equation}
The nonlinear model of the type (6) appears due to the vacuum
polarization of arbitrary spin particles \cite{Kruglov}. The
propagation of a linearly polarized laser light in the external
transverse magnetic field, in the effective theory (6), was
investigated in \cite{Kruglov1}. If $\beta\neq\gamma$, the model
(6) leads to birefringence: the phase velocities of light depend
on polarizations in the external electromagnetic fields. There is
no birefringence in BI electrodynamics. The Heisenberg$-$Euler
Lagrangian \cite{Heisenberg}, \cite{Schwinger}, \cite{Adler}
corresponding to the one-loop corrections to classical
electrodynamics follows form (6), (7) at $B\rightarrow \infty$. Therefore, the case
$\beta\neq\gamma$ is natural due to quantum corrections of
Maxwell's electrodynamics. The case of the finiteness of $\beta$ and
vanishing $\gamma^{-2}$ ($\gamma\rightarrow \infty$) in equation (6)
(ignoring relations (7)), was considered \cite{Kerner} in the framework
of the Kaluza$-$Klein theory (in five dimensions) with the additional Gauss$-$Bonnet term.
Now we can introduce the generalized BI Lagrangian density
\begin{equation}
{\cal L} =
\beta^2\left(1-\sqrt{1+2\beta^{-2}S-\beta^{-2}\gamma^{-2}P^2}\right),
 \label{8}
\end{equation}
which leads to equation (6) at small values of $\beta^{-2}S$, $\beta^{-2}\gamma^{-2}P^2$.
Thus, the model with Lagrangian density (8) with two different parameters $\beta$ and
$\gamma$ can be considered as the natural generalization of BI
electrodynamics. We treat this non-linear model
as the effective electrodynamics theory for
strong electromagnetic fields. This model allows us to consider
particular cases. Of course our discussion
of motivating Lagrangian density (8) is not strict enough and is heuristic.
One may consider the model based on equation (8) as a convenient parametrization
which allows us to analyze different models discussed in literature.
Anyway, a model (8) in our opinion has a definite theoretical interest.

\section{The field equations}

Introducing, as usual, the four-vector potential $A_{\mu}$:
$F_{\mu\nu}=\partial_\mu A_\nu -\partial_\nu A_\mu$, using Eq.(8),
we find the equations of motion from the Euler-Lagrange equations
($\partial_\mu\partial{\cal L}/\partial(\partial_\mu
A_\nu)-\partial{\cal L}/\partial A_\nu =0$):
\begin{equation}
\partial_\mu\left[\frac{1}{{\cal R}}\left(F_{\mu\nu}-\frac{P}{\gamma^2}
\widetilde{F}_{\mu\nu}\right)\right]=0, \label{9}
\end{equation}
where
\begin{equation}
{\cal R}=\sqrt{1+2\beta^{-2}S-\beta^{-2}\gamma^{-2}P^2}. \label{10}
\end{equation}
Other equations follow from the Bianchi identity:
\begin{equation}
\partial_\mu \widetilde{F}_{\mu\nu}=0.
\label{11}
\end{equation}
Equations (9),(11) may be cast into the form of nonlinear Maxwell's
equations. Introducing the electric displacement field
$\textbf{D}=\partial{\cal L}/\partial \textbf{E}$, one obtains
from Eq.(8)
\begin{equation}
\textbf{D}=\frac{1}{{\cal R}}\left(
\textbf{E}+\frac{1}{\gamma^2}\textbf{B}(\textbf{E}\cdot\textbf{B})\right).
\label{12}
\end{equation}
The magnetic field is given by $\textbf{H}=-\partial{\cal
L}/\partial \textbf{B}$:
\begin{equation}
\textbf{H}= \frac{1}{{\cal R}}\left(
\textbf{B}-\frac{1}{\gamma^2}\textbf{E}(\textbf{E}\cdot\textbf{B})\right),
\label{13}
\end{equation}
where $E_j=iF_{j4}$, $B_j=(1/2)\varepsilon_{jik}F_{ik}$ ($\varepsilon_{123}=1$). Eq.(9)
can be represented in the terms of fields $\textbf{D}$ and
$\textbf{H}$:
\begin{equation}
\nabla\cdot \textbf{D}= 0,~~~~ \frac{\partial\textbf{D}}{\partial
t}-\nabla\times\textbf{H}=0. \label{14}
\end{equation}
The second pair of Maxwell's equation follows from equation (11):
\begin{equation}
\nabla\cdot \textbf{B}= 0,~~~~ \frac{\partial\textbf{B}}{\partial
t}+\nabla\times\textbf{E}=0. \label{15}
\end{equation}
From equations (12),(13), we obtain the electric permittivity
$\varepsilon_{ik}$ and the inverse magnetic permeability
$(\mu)^{-1}_{ik}$ tensors of the vacuum
\begin{equation}
\varepsilon_{ik}=\frac{1}{{\cal R}}\left(\delta_{ik}+
\frac{1}{\gamma^2}B_iB_k\right),~~~~
(\mu)^{-1}_{ik}=\frac{1}{{\cal R}}\left(\delta_{ik}-
\frac{1}{\gamma^2}E_iE_k\right),
  \label{16}
\end{equation}
with the connections
\begin{equation}
 D_i=\varepsilon_{ik}E_k ,~~~~B_i=\mu_{ik}H_k. \label{17}
\end{equation}
Thus, the generalized BI electrodynamics equations with two
parameters were rewritten in the form of Maxwell's equations
(14),(15) with the nonlinear links (17) (because of (16)).

Equations (16) show that the vacuum possesses the complicated anisotropic
properties. Therefore, it is useful to find eigenvalues of the
electric permittivity and magnetic permeability tensors. From
(16), one finds the ``minimal" polynomials of matrices
$\varepsilon=(\varepsilon_{ik})$ and $\mu^{-1}=((\mu^{-1})_{ik})$:
\[
\left[\varepsilon-\frac{1}{{\cal
R}}\left(1+\frac{\textbf{B}^2}{\gamma^2}\right)\right]\left(\varepsilon-\frac{1}{{\cal
R}}\right)=0,
\]
\vspace{-7mm}
\begin{equation} \label{18}
\end{equation}
\vspace{-7mm}
\[
\left[\mu^{-1}-\frac{1}{{\cal
R}}\left(1-\frac{\textbf{E}^2}{\gamma^2}\right)\right]\left(\mu^{-1}-\frac{1}{{\cal
R}}\right)=0.
\]
Equations (18) allow us to obtain eigenvalues
$\lambda_{1,2}(\varepsilon,\mu^{-1})$ and inverse matrices
$\varepsilon^{-1}$, $\mu$:
\[
\lambda_{1}(\varepsilon)=\frac{1}{{\cal
R}}\left(1+\frac{\textbf{B}^2}{\gamma^2}\right),~~~~\lambda_{2}(\varepsilon)=\frac{1}{{\cal
R}},
\]
\vspace{-7mm}
\begin{equation} \label{19}
\end{equation}
\vspace{-7mm}
\[
\lambda_{1}(\mu^{-1})=\frac{1}{{\cal
R}}\left(1-\frac{\textbf{E}^2}{\gamma^2}\right),~~~~\lambda_{2}(\mu^{-1})=\frac{1}{{\cal
R}},
\]
\[
\left(\varepsilon^{-1}\right)_{ik}={\cal
R}\left(\delta_{ik}-\frac{\gamma^{-2}B_iB_k}{1+\gamma^{-2}\textbf{B}^2}
\right),
\]
\vspace{-7mm}
\begin{equation} \label{20}
\end{equation}
\vspace{-7mm}
\[
\mu_{ik}={\cal
R}\left(\delta_{ik}+\frac{\gamma^{-2}E_iE_k}{1-\gamma^{-2}\textbf{E}^2}\right).
\]
As in BI electrostatics (at $\textbf{B}=\textbf{H}=0$), we write
one of equations (14) with the point source
\begin{equation}
\nabla\cdot \textbf{D}_0=e\delta(\textbf{r}) \label{21}
\end{equation}
possessing the solution
\begin{equation}
\textbf{D}_0=\frac{e}{4\pi r^3}\textbf{r}. \label{22}
\end{equation}
Then from (16),(17), one obtains the electric field
\begin{equation}
\textbf{E}_0=\frac{\textbf{D}_0}{\sqrt{1+\beta^{-2}D_0^2}}=
\frac{e\textbf{r}}{4\pi r\sqrt{r^4+r_0^4}}, \label{23}
\end{equation}
where the elementary length is $r_0=\sqrt{\mid e\mid/4\pi} \beta$.
The electric field of a point-like charge is not singular
resulting to the finiteness of energy contrarily to linear
electrodynamics. Thus, the parameter $\gamma$ introduced does not
enter electrostatics equations and all equations remain the same
as in BI electrodynamics with the replacement $B\rightarrow
\beta$. It follows from equation (23) that the maximum field strength is
$E_{max}=\lim_{r\rightarrow 0}E_0=\beta$. The lower bound on the
BI parameter $B$ \cite {Jackson} is given by $B\geq 10^{20}$ V/m.

From equations (12),(13), we obtain the relation
\begin{equation}
\textbf{D}\cdot\textbf{H}=\textbf{E}\cdot\textbf{B}
\frac{1+2\gamma^{-2}S-\gamma^{-4}P^2}{1+2\beta^{-2}S-\beta^{-2}\gamma^{-2}P^2}.
\label{24}
\end{equation}
According to the criterion of  \cite{Gibbons1}, the nonlinear
electrodynamics possesses the duality symmetry if
$\textbf{D}\cdot\textbf{H}=\textbf{E}\cdot\textbf{B}$. It follows
from (24) that the duality symmetry is broken in the model
considered if $\beta\neq\gamma$. In BI electrodynamics
$\beta=\gamma\equiv B$ and the duality symmetry is recovered. In
QED due to quantum corrections $4a\neq b$ ($\beta\neq\gamma$, see equations (5),(7)),
and the duality symmetry is broken.

\section{Propagation of waves in the background magnetic field}

Now we analyze the propagation of the plane electromagnetic wave
($\textbf{e},\textbf{b}$) traveling in $z$-direction and
perpendicular to the external constant and uniform magnetic
induction field $\overline{\textbf{B}}=(\overline{B},0,0)$. We
consider the Lagrangian (8) without the approximation of weakness of
external fields. The electromagnetic fields are the sum of the
light field and the background magnetic field,
$\textbf{E}=\textbf{e}$,
$\textbf{B}=\textbf{b}+\overline{\textbf{B}}$. It is implied that
the wave fields are much weaker than external magnetic induction
field. Replacing the decomposition of fields in equation (8), and
neglecting the higher order in light fields, one obtains the
Lagrangian
\begin{equation}
{\cal L}\left(\textbf{e},\textbf{b}+\overline{\textbf{B}}\right) =
\beta^2\left(1-\sqrt{1+\frac{\left(\textbf{b}+\overline{\textbf{B}}\right)^2-\textbf{e}^2}
{\beta^{2}}-\frac{\left(\textbf{e}\cdot\textbf{B}\right)^2}{\beta^{2}\gamma^{2}}}\right).
 \label{25}
\end{equation}
We left in (25) only quadratic terms in $\textbf{e}$ and
$\textbf{b}$. One can find from equation (25) the displacement and
induction magnetic vectors in the background fields
\begin{equation}
d_i=\frac{\partial {\cal L}}{\partial
e_i}=\frac{1}{\kappa}\left(\delta_{ij}+\frac{1}{\gamma^2}
\overline{B}_i\overline{B}_j\right)e_j, \label{26}
\end{equation}
\begin{equation}
h_i=- \frac{\partial {\cal L}}{\partial
b_i}=\frac{1}{\kappa}\left(\delta_{ij}-\frac{1}{\kappa^2\beta^2}
\overline{B}_i\overline{B}_j\right)b_j, \label{27}
\end{equation}
where
\begin{equation}
\kappa=\sqrt{1+\frac{\overline{\textbf{B}}^2}{\beta^{2}}},
\label{28}
\end{equation}
and only linear terms in $\textbf{e}$ and $\textbf{b}$ are left in
(26),(27). It should be noted that we do not imply the
smallness of the background magnetic fields. Thus, the unitless
parameters $\overline{\textbf{B}}^2/\beta^{2}$,
$\overline{\textbf{B}}^2/\gamma^{2}$ can be arbitrary. From
equations (26),(27), using the relations $d_i=\varepsilon_{ij}e_j$,
$h_i=(\mu^{-1})_{ij}b_j$, we obtain the polarization tensors in
the external magnetic field
\begin{equation}
\varepsilon_{ij}=\frac{1}{\kappa}\left(\delta_{ij}+\frac{1}{\gamma^2}
\overline{B}_i\overline{B}_j\right),~~~~\left(\mu^{-1}\right)_{ij}=
\frac{1}{\kappa}\left(\delta_{ij}-\frac{1}{\kappa^2\beta^2}
\overline{B}_i\overline{B}_j\right).\label{29}
\end{equation}
From Maxwell equations, one finds the equation for the electric
wave field as follows \cite{Kruglov1}:
\begin{equation}
\left[\textbf{k}^2\left(\mu^{-1}\right)_{bi}+k_a\left(\mu^{-1}\right)_{al}k_l\delta_{ib}-
\textbf{k}^2\left(\mu^{-1}\right)_{aa}\delta_{ib}-k_l\left(\mu^{-1}\right)_{bl}k_i+
\omega^2\varepsilon_{ib}\right]e_b=0 . \label{30}
\end{equation}
It follows from homogeneous equation (30) that nontrivial solutions
exist when the determinant of the matrix is zero. Replacing
(29) into (30), we obtain the corresponding matrix
\begin{equation}
\Lambda_{ij}=\left(1-n^2 +n^2
\frac{\overline{\textbf{B}}^2}{\kappa^2\beta^{2}}\right)\delta_{ij}
+\left(\frac{1}{\gamma^2}-\frac{n^2}{\kappa^2\beta^{2}}\right)\overline{B}_i\overline{B}_j,
\label{31}
\end{equation}
where the index of refraction is $n=k/\omega$. With
the help of the method of \cite{Kruglov1}, one finds the
eigenvalues of the matrix (31):
\begin{equation}
\lambda_1=1-n^2\left(1 -
\frac{\overline{\textbf{B}}^2}{\kappa^2\beta^{2}}\right),~~~~
\lambda_2=1-n^2+\frac{\overline{\textbf{B}}^2}{\gamma^{2}}.
\label{32}
\end{equation}
From equation (32), with the help of (28), we obtain the indexes of
refraction for two modes corresponding to $\lambda_1=0$ and
$\lambda_2=0$:
\begin{equation}
n_\bot=\sqrt{1+\frac{\overline{\textbf{B}}^2}{\beta^{2}}},~~~~
n_\|=\sqrt{1+\frac{\overline{\textbf{B}}^2}{\gamma^{2}}}.
\label{33}
\end{equation}
Thus, the electromagnetic waves with different polarizations have
different velocities $v_\bot=n_\bot^{-1}$, $v_\|=n_\|^{-1}$, and
there is the effect of vacuum birefringence if $\beta\neq\gamma$.
Let the polarization vector at $z=0$ is
$\textbf{e}|_{z=0}=E_0(\cos\theta, \sin\theta)\exp(-i\omega t)$,
where $\theta$ is the angle between the polarization vector
$\textbf{e}$ and the external magnetic induction field
$\overline{\textbf{B}}$. Then the linearly polarized wave
traveling the distance $L$ becomes the elliptically polarized
wave. According to \cite{Kruglov1}, ellipticity (the ratio of
minor to major axis of the ellipse) is given by
\begin{equation}
\Psi=\frac{1}{2}\left(n_\bot-n_\|\right)\omega L\sin2\theta ,
\label{34}
\end{equation}
where $\omega=2\pi/\lambda$, and $\lambda$ is a wave length. In BI
electrodynamics $\beta=\gamma$ ($n_\bot=n_\|$) and vacuum birefringence vanishes.

In the case of the smallness of the parameters
$\overline{\textbf{B}}^2/\beta^{2}$, $\overline{\textbf{B}}^2/\gamma^{2}$, the indexes of refraction
(33) become
\begin{equation}
n_\bot\simeq 1+\frac{1}{2\beta^2}\overline{B}^2,
~~~~n_\|\simeq 1+\frac{1}{2\gamma^2}\overline{B}^2 .\label{35}
\end{equation}
Replacing equation (35) into (34), one obtains the result of
\cite{Kruglov1} corresponding to the Lagrangian (6).
Let us estimate the upper bounds on the value
\begin{equation}
\Delta=\frac{1}{\beta^2}-\frac{1}{\gamma^2}
\label{36}
\end{equation}
with the help of polarization data for ellipticity $\Psi$ of BRST \cite{BRST}
and PVLAS \cite{PVLAS} Collaborations. The experimental apparatus of BRST Collaboration consisted of
a magnetic-field region and the ellipsometer, where the resulting polarization change was measured.
The magnetic field was supplied by two superconducting dipole magnets, and the length of the two magnets was 8.8 m
(1 m=$5.1\times10^6$ eV$^{-1}$).
They studied the propagation of a laser beam with the wavelength $\lambda=514.5$ nm through a transverse magnetic field of 3.25 T ($\overline{B}=3.25$ T, 1 T=195.5 eV$^2$ in Heaviside$-$Lorentz's units), and searched for light scalar and/or pseudoscalar particles that couple to two photons. The input polarizer was set at $45^\circ$ ($\theta=45^\circ$) to the direction of the magnetic field. The number of reflections was $N$ so that $NL$ is the total optical path length.
From BRST \cite{BRST} data, one obtains
\[
N=578, ~~~~ \Psi=40~nrad,~~~~ \Delta=6.38\times10^{-24}~eV^{-4},
\]
\[
 N=34,~~~~ \Psi=1.6~nrad,~~~~ \Delta=4.34\times10^{-24}~eV^{-4}.
\]

In the PVLAS experiment, the wavelength of $\lambda=1064$ nm was used with the magnetic field strengths of $ \overline{B}= 2.3$ T. The setup consisted of a sensitive ellipsometer detected changes in the polarization state of light propagating through a $L=1$ m long magnetic field region in vacuum. It was based on a high finesse Fabry-Perot cavity and a superconducting rotating dipole magnet. The results from measurements did not confirm the presence of a rotation signal and excluded an ellipticity signal at 2.3 T. To check a presence of fringe field effects, two measurements in vacuum were performed with the apparatus in the configuration at $ \overline{B}= 2.3$ ~ T field (there was no a fringe field), and at a 5 T field (there was a fringe field). Thus, at the 2.3 T measurements, no visible signal peak was observed both in rotation and in ellipticity. It was concluded in \cite{PVLAS} that the rotation measurements at the field intensity of 5 T indicated that the rotation signal reported in previous publications was due to an instrumental artifact. It should be noted that the rotation of the magnetic field in the PVLAS experiment does not effect on the value of vacuum birefringence within QED calculations \cite{Adler}, \cite{Biswas}.
The limiting observed background value for ellipticity is $\Psi<0.31$ prad/pass with
$N=45000$ passes in the interaction region. Using the value  $ \overline{B}= 2.3$ T, $L=1$ m, and $\theta=\pi$/4 of PVLAS setup, we obtain the upper bound on the value of the parameter $\Delta$ (36):
\[
\Delta<1\times10^{-24}~eV^{-4}.
 \]
We note that new results exclude the particle interpretation of the
previous PVLAS results as due to a spin zero boson.

\section{The energy-momentum tensor and dilatation current}

With the help of the Lagrangian (8), we find the conserved
canonical energy-momentum tensor $T_{\mu\nu}^{c}=(\partial_\nu
A_\alpha)\partial{\cal L}/\partial(\partial_\mu
A_\alpha)-\delta_{\mu\nu}{\cal L}$:
\begin{equation}
T_{\mu\nu}^{c}=(\partial_\nu A_\alpha)\frac{1}{{\cal
R}}\left(\frac{P}{\gamma^2}
\widetilde{F}_{\mu\alpha}-F_{\mu\alpha}\right)-\delta_{\mu\nu}{\cal
L}, \label{37}
\end{equation}
and $\partial_\mu T^c_{\mu\nu}=0$. The tensor (37) is not
symmetric and gauge-invariant tensor. We can obtain the symmetric
Belinfante tensor using the relation \cite{Coleman}:
\begin{equation}
T_{\mu\nu}^{B}=T_{\mu\nu}^{c}+\partial_\beta X_{\beta\mu\nu},
\label{38}
\end{equation}
where
\begin{equation}
X_{\beta\mu\nu}=\frac{1}{2}\left[\Pi_{\beta\sigma}\left(\Sigma_{\mu\nu}\right)_{\sigma\rho}
-\Pi_{\mu\sigma}\left(\Sigma_{\beta\nu}\right)_{\sigma\rho}-
\Pi_{\nu\sigma}\left(\Sigma_{\beta\mu}\right)_{\sigma\rho}\right]A_\rho,
\label{39}
\end{equation}
\begin{equation}
\Pi_{\mu\sigma}=\frac{\partial{\cal L}}{\partial(\partial_\mu
A_\sigma)}=\frac{1}{{\cal R}}\left(\frac{P}{\gamma^2}
\widetilde{F}_{\mu\sigma}-F_{\mu\sigma}\right). \label{40}
\end{equation}
The tensor $X_{\beta\mu\nu}$ is antisymmetric in indexes
$\beta$ and $\mu$, and therefore $\partial_\mu\partial_\beta
X_{\beta\mu\nu}=0$. As a result $\partial_\mu T^B_{\mu\nu}=\partial_\mu T^c_{\mu\nu}=0$ and the symmetric
Belinfante tensor is also conserved. The generators of the Lorentz transformations
$\Sigma_{\mu\alpha}$ have the matrix elements:
\begin{equation}
\left(\Sigma_{\mu\alpha}\right)_{\sigma\rho}=\delta_{\mu\sigma}\delta_{\alpha\rho}
-\delta_{\alpha\sigma}\delta_{\mu\rho}. \label{41}
\end{equation}
With the help of equations (39),(40), one finds
\begin{equation}
\partial_\beta X_{\beta\mu\nu}=\Pi_{\beta\mu}\partial_\beta A_\nu.
\label{42}
\end{equation}
It follows from equation of motion (9) that
$\partial_\mu\Pi_{\mu\nu}=0$, and one can verify that the equation
$\partial_\mu\partial_\beta X_{\beta\mu\nu}=0$ holds. Using
equations (40),(42), the conserved Belinfante tensor (38) becomes
symmetric and gauge-invariant tensor:
\begin{equation}
T_{\mu\nu}^{B}=\frac{F_{\nu\alpha}}{{\cal
R}}\left(\frac{P}{\gamma^2}
\widetilde{F}_{\mu\alpha}-F_{\mu\alpha}\right)-\delta_{\mu\nu}{\cal
L}. \label{43}
\end{equation}
The symmetric energy-momentum tensor (43) also can be obtained by
varying the action $S=\int d^4x{\cal L}$ (after formally using the
curve space-time) on the symmetric metric tensor $g^{\mu\nu}$. The
trace of the energy-momentum tensor (43) is not zero:
\begin{equation}
T_{\mu\mu}^{B}=4\beta^2\left(\frac{1+\beta^{-2}S}{{\cal
R}}-1\right). \label{44}
\end{equation}
According to \cite{Coleman}, we define the modified dilatation
current
\begin{equation}
D_{\mu}^{B}=x_\alpha T_{\mu\alpha}^{B}+V_\mu, \label{45}
\end{equation}
where the field-virial $V_\mu$ is given by
\begin{equation}
V_\mu=\Pi_{\alpha\beta}\left[\delta_{\alpha\mu}\delta_{\beta\rho}
-\left(\Sigma_{\alpha\mu}\right)_{\beta\rho}\right]A_\rho=0.
\label{46}
\end{equation}
As the field-virial vanishes in our case, the modified dilatation
current (45) becomes $D_{\mu}^{B}=x_\alpha T_{\mu\alpha}^{B}$.
Then the four-divergence of dilatation current is equal to the
trace of the energy-momentum tensor (44):
\begin{equation}
\partial_\mu D_{\mu}^{B}=T_{\mu\mu}^B. \label{47}
\end{equation}
As a result, the dilatation (scale) symmetry is broken because of
the presence of the parameter $\beta$ with the dimension of the
field strength. Thus, there is a difference compared to
linear Maxwell electrodynamics: Maxwell equations are scale
invariant. The conformal symmetry, which includes the
one-parameter dilatation group of symmetry, is also broken
\cite{Coleman}.

\section{Conclusion}

We have formulated the generalized BI electrodynamics with two
parameters $\beta$ and $\gamma$. The model includes particular cases: BI
electrodynamics if $\beta=\gamma$, linear Maxwell electrodynamics with
quantum non-linear Heisenberg$-$Euler
corrections, electrodynamics with one loop corrections due
to vacuum polarization of arbitrary spin particles and others. We
have obtained induced ellipticity in the background magnetic field
due to the effect of vacuum birefringence. The upper bounds on the parameter
$\Delta=1/\beta^2-1/\gamma^2$ are obtained from the experimental data of BRST \cite{BRST}
and PVLAS \cite{PVLAS} Collaborations. For the case of BI electrodynamics,
phase velocities of different polarizations of
the light beam are equal, and the effect of vacuum birefringence
vanishes. It should be noted that the effect of vacuum birefringence is
now of the experimental interest \cite{BRST}, \cite{PVLAS}, \cite{QA}.
The canonical and symmetrical Belinfante energy-momentum tensors, and the dilatation current
have been obtained. It was demonstrated that the dilatation current is not conserved due
to the non-zero trace of the energy-momentum tensor. Thus, the
scale symmetry is broken because of the presence of the
dimensional parameter $\beta$. If $\beta\neq\gamma$ the dual
symmetry is also broken.


\begin{thebibliography}{99}

\bibitem{Tseytlin} E.S.Fradkin and A.A.Tseytlin, Phys. Lett. \textbf{163B}, 123 (1985).


\bibitem{Born} M.Born, Proc. R. Soc. London \textbf{A143}, 410 (1934).

\bibitem{Infeld} M.Born and M.Infeld, Proc. R. Soc. London \textbf{A144}, 425 (1934).

\bibitem{Plebanski} J.Pleba\'{n}ski, Lectures on
non-linear electrodynamics, at Nordita, Copenhagen, 1968 (1970).

\bibitem{Boillat} G.Boillat, J. Math. Phys. \textbf{11}, 941 (1970).

\bibitem{Gibbons} G.W. Gibbons, Grav. Cosm., \textbf{8}, 2 (2002) [arXiv:hep-th/0104015].

\bibitem{Aiello} M.Aiello,. G.R.Bengochea, and R.Ferraro, Phys. Lett. \textbf{A361}, 9 
(2007) [arXiv:hep-th/0607072].


\bibitem{Kruglov} S.I.Kruglov, Ann. Phys. \textbf{293}, 228
(2001) [arXiv:hep-th/0110061].

\bibitem{Kruglov1} S.I.Kruglov, Phys. Rev. \textbf{D 75}, 117301 (2007).

\bibitem{Heisenberg}  W.Heisenberg and H.Euler, Z. Physik \textbf{98}, 714 (1936).

\bibitem{Schwinger} J.Schwinger, Phys. Rev. D\textbf{82}, 664 (1951).

\bibitem{Adler} S.L.Adler, Ann. Phys. \textbf{67}, 599 (1971).

\bibitem{Kerner} J.P.S.Lemos, R.Kerner, Grav. Cosm., \textbf{6}, 49
(2000) [arXiv:hep-th/9907187].

\bibitem{Jackson} J.D.Jackson, Classical electrodynamics (Jhon
Wiley and Sons, N.Y. 1999).

\bibitem{Gibbons1} G.W.Gibbons, D.Rasheed, Nucl. Phys., \textbf{B 454}, 185
(1995) [arXiv:hep-th/9506035].

\bibitem{BRST} R.Cameron et al. [BRST Collaboration], Phys. Rev. \textbf{D 47}, 3707 (1993).

\bibitem{PVLAS} E.Zavattini et al., [PVLAS Collaboration], Phys. Rev. \textbf{D 77}, 032006 (2008)
[arXiv:0706.3419 [hep-ex]].

\bibitem{Adler} S.L.Adler, J. Phys. \textbf{A40} (2007), F143 [arXiv:hep-ph/0611267].

\bibitem{Biswas} S.Biswas and K.Melnikov, Phys. Rev. \textbf{D75} (2007),
053003 [arXiv:hep-ph/0611345].


\bibitem{Coleman} S.Coleman and R.Jackiw, Ann. Phys. \textbf{67}, 552 (1971).


\bibitem{QA} S.J.Chen, H.H.Mei and W.T.Ni [Q\&A Collaboration] Mod. Phys. Lett.
\textbf{A 22}, 2815 (2007) [arXiv:hep-ex/0611050].

\end{thebibliography}
\end{document}